\documentclass[a4paper,11pt]{article}

\usepackage{jheppub}

\usepackage{amsfonts}
\usepackage{amsmath}
\usepackage{amssymb}
\usepackage[titletoc,toc,title]{appendix}
\usepackage{braket}
\usepackage{cancel}
\usepackage[font={footnotesize,it}]{caption}
\usepackage{color}
\usepackage{comment}
\usepackage{default}
\usepackage{enumitem}
\usepackage{epsfig}
\usepackage{epstopdf}
\usepackage{float}
\usepackage{graphicx}
\usepackage{hyperref}
\usepackage{cleveref}
\usepackage[utf8]{inputenc}
\usepackage[numbers]{natbib}
\usepackage{mathtools}
\usepackage{ragged2e}
\usepackage{subcaption}
\usepackage{tikz}

\hypersetup{
citecolor=red,
colorlinks=true,
filecolor=red,
linkcolor=blue,
linktocpage=true,
urlcolor=blue
}

{\makeatletter \g@addto@macro\bfseries{\boldmath} \makeatother}

\usetikzlibrary{shapes}
\begin{document}

\title{Time Evolution of Entanglement Negativity from Black Hole Interiors}

\author[a]{Vinay Malvimat,}
\author[b]{Sayid Mondal,}
\author[b]{and Gautam Sengupta}

\affiliation[a]{
Indian Institute of Science Education and Research,\\
Homi Bhabha Rd, Pashan, Pune 411 008, India
\bigskip
}

\affiliation[b]{
Department of Physics,\\
Indian Institute of Technology,\\
Kanpur 208 016, India
\bigskip
}

\emailAdd{vinaymmp@gmail.com}
\emailAdd{sayidphy@iitk.ac.in}
\emailAdd{sengupta@iitk.ac.in}

\abstract{
\noindent
We investigate the time evolution of entanglement negativity following a global quench for mixed state configurations of two disjoint and adjacent intervals in a ($1+1$)-dimensional conformal field theory ($CFT_{1+1}$) dual to the eternal black hole sliced in half by an end of the world brane, through the $AdS_3/CFT_2$ correspondence. To this end we obtain the time evolution of the holographic entanglement negativity for such mixed states from a dual bulk eternal black hole geometry and elucidate the relevant geodesic structures. The holographic entanglement negativity for such mixed states, following a global quench is described by half of the results for the eternal black hole. Significantly our results exactly match with the corresponding $CFT_{1+1}$ computations.
}
\arxivnumber{1812.04424}

\maketitle


\section{Introduction}
\label{sec1}
\justify

Over the last few years quantum entanglement has emerged as a central issue in the study of diverse physical phenomena ranging from quantum many-body systems in out-of-equilibrium to the process of black hole formation and the information loss paradox. The entanglement for a bipartite pure state is characterized by the entanglement entropy which is the von Neumann entropy of the reduced density matrix. In $(1+1)$-dimensional conformal field theories ($CFT_{1+1}$) the entanglement entropy may be computed through a replica technique as described in \cite{Calabrese:2004eu,Calabrese:2009qy}.
However entanglement entropy fails to characterize mixed state entanglement as it
typically involves correlations irrelevant to the entanglement of the specific mixed state. This subtle issue was addressed in quantum information theory by Vidal and Werner in \cite{PhysRevA.65.032314} where the authors proposed a computable measure termed as {\em entanglement negativity} which characterized the upper bound on the distillable entanglement for the mixed state\footnote {It was demonstrated by Plenio in \cite{Plenio:2005cwa} that this entanglement measure was not convex but was an entanglement monotone under local operations and classical communication (LOCC).}. The entanglement negativity is defined as the logarithm of the trace norm  of the partially transposed density matrix with respect to a subsystem.  In a series of interesting communications the authors in \cite{Calabrese:2012ew,Calabrese:2012nk,Calabrese:2014yza} described the computation of the entanglement negativity for bipartite mixed states in a $CFT_{1+1}$ through a suitable replica technique.

A holographic characterization of the entanglement entropy in $CFT$s was proposed by Ryu and Takayanagi \cite{Ryu:2006bv,Ryu:2006ef} through the $AdS/CFT$ correspondence. According to their conjecture the universal part of the entanglement entropy of a subsystem in a dual $d$-dimensional conformal field theories ($CFT_d$) was proportional to the area of a co-dimension two static minimal surface in the bulk $AdS_{d+1}$ geometry homologous to the subsystem. Naturally, this led to intense research activity in various aspects of entanglement related issues in holographic $CFT$s \cite{Takayanagi:2012kg,Nishioka:2009un,Nishioka:2018khk,e12112244,Blanco:2013joa,Fischler2013,Fischler:2012uv,Chaturvedi:2016kbk} (and references therein). In a series of interesting communications, a proof of this conjecture was developed from a bulk perspective, initially in the context of $AdS_3/CFT_2$ and later generalized to the $AdS_{d+1}/CFT_d$ scenario \cite{Fursaev:2006ih,Headrick:2010zt,Faulkner:2013yia,Casini:2011kv,Lewkowycz:2013nqa}. A covariant extension of the RT conjecture was proposed for the entanglement entropy of a subsystem in a $CFT_d$ dual to a non static bulk $AdS_{d+1}$ geometry in \cite{Hubeny:2007xt} and the corresponding proof in \cite {Dong:2016hjy}.

The above developments naturally lead to the interesting question of a holographic characterization of the entanglement negativity for bipartite pure and mixed states in dual $CFT$s \cite{Rangamani:2014ywa,Perlmutter:2015vma}. In the recent past two of the present authors (VM and GS) in a collaboration, proposed a holographic construction for the entanglement negativity of bipartite pure and mixed states in a dual $CFT_{1+1}$ through $AdS_3/CFT_2$ framework and its covariant extension \cite{Chaturvedi:2016rcn,Chaturvedi:2016opa}. This construction was rigorously substantiated through a large central charge analysis employing the monodromy technique in \cite{Malvimat:2017yaj}, although a bulk proof for this conjecture along the lines of \cite{Faulkner:2013yia} is an outstanding issue. Subsequently, their proposal was also extended to higher dimensions in the context of the $AdS_{d+1}/CFT_d$ scenario in \cite{Chaturvedi:2016rft}. It should be mentioned here that a proof of the higher dimensional extension for the holographic entanglement negativity conjecture described above from a bulk perspective along the lines of \cite{Lewkowycz:2013nqa} is also an outstanding non trivial open issue. Subsequent to this in \cite{Jain:2017aqk,Jain:2017uhe}, a holographic characterization for the entanglement negativity of a mixed state of adjacent intervals in $CFT_{1+1}$s dual to bulk $AdS_3$ geometries, and its covariant generalization was proposed. A higher dimensional generalization of the above holographic construction for such mixed states of adjacent subsystems in a $CFT_d$ dual to bulk $AdS_{d+1}$ geometry, was subsequently advanced in \cite{Jain:2017xsu,Jain:2018bai}. Furthermore very recently a holographic construction for the entanglement negativity of mixed states of two disjoint intervals in a dual $CFT_{1+1}$ and its covariant extension has been developed in \cite{Malvimat:2018txq,Malvimat:2018ood}. 

In a different context the study of out of equilibrium quantum systems to elucidate their dynamical evolution has witnessed a strong surge of interest in recent times \cite{Calabrese:2005in,Calabrese:2009qy,calabrese2016quantum,Wen:2018svb,Wen:2015qwa}. One of the processes through which a quantum system may be placed in an out of equilibrium configuration is a quantum quench. In a global quench scenario a quantum system is initially prepared in the ground state of a translationally invariant Hamiltonian $H_0$ and is then allowed to evolve unitarily with respect to another Hamiltonian $H$, where $H$ and $H_0$ are related to each other through an experimentally tunable parameter. In a 
$CFT_{1+1}$ the resulting state after a global quench may be regarded as a boundary state (B-state) as it is translational and conformal invariant below the energy scale of the global quench. The corresponding time evolution of the entanglement entropy after a global quench for a subsystem maybe computed through the replica technique \cite{Calabrese:2005in,Calabrese:2009qy,calabrese2016quantum} in a $CFT_{1+1}$. It was shown that in the spacetime scaling limit\footnote {In the spacetime scaling limit the time $t$ and the interval length $\ell$ are much larger than the microscopic length and time scale in the theory \cite{Coser:2014gsa,PhysRevA.78.010306,PhysRevLett.106.227203,Calabrese:2009qy}. } the entanglement entropy grows linearly for early times and is rendered extensive beyond a certain value of time. It is possible to describe such a time evolution of the entanglement following a global quench, through a quasi-particle scenario for the entanglement propagation \cite{Calabrese:2005in,calabrese2016quantum,Calabrese:2007rg,Calabrese:2006rx}. In this instance the initial state which is at a higher energy relative to the ground state of the post-quench Hamiltonian, acts as a source of quasi-particle excitations. 

Following the developments described above the corresponding time evolution of the entanglement negativity for mixed states of two adjacent and disjoint intervals in a $CFT_{1+1}$ following a global quench was investigated in \cite{Coser:2014gsa}. Very interestingly they demonstrated that similar to the entanglement entropy, the time evolution of entanglement negativity also admits of a quasi-particle explanation for the mixed state entanglement propagation. However it was observed that the initial increase in the entanglement negativity occurs slightly after the time predicted by the quasi-particle picture as a consequence of the late birth of entanglement in a lattice system which disappears in the continuum limit \cite{Coser:2014gsa}.

On the other hand the time evolution of holographic entanglement entropy after a global quench in a $CFT_{1+1}$ dual to a bulk Vaidya-AdS geometry describing black hole formation
has been extensively studied in the literature \cite{Hubeny:2007xt,AbajoArrastia:2010yt,Aparicio:2011zy,Albash:2010mv,Balasubramanian:2010ce,Balasubramanian:2011ur,Allais:2011ys,Liu:2013iza,Liu:2013qca,Ugajin:2013xxa,Liu:2018crr,Anous:2017tza}. Although as discussed in \cite{Hartman:2013qma}, the precise holographic dual for the global quench scenario where the initial state is a time dependent pure boundary state (B-state) in a $CFT_{1+1}$ \cite{Calabrese:2005in}, is the eternal black hole geometry sliced in half by an end-of-the-world (ETW) brane or simply a {\it single sided black hole}\footnote{Recently in \cite{Cooper:2018cmb}, a somewhat similar bulk configuration is considered where the end-of-the-world (ETW) brane is extended into the second asymptotic region.}. It is well known from \cite{Israel:1976ur,Maldacena:2001kr,Hartman:2013qma} that the eternal black hole is dual to the thermofield double state in a $CFT$. These involve two-sided Penrose diagram and are static under time evolution as the time directions are taken to be opposite for the two sides. However, the time dependence may be introduced for this geometry by considering forward time evolution for both the exterior regions of the Penrose diagram. 

The computation of the corresponding holographic entanglement entropy for a subsystem $A$ in a $CFT$ for the thermofield double state dual to the bulk eternal black hole geometry requires the consideration of the subsystem in question on both sides of the Penrose diagram. The entanglement entropy may then be computed from the area of the extremal surface anchored on both the subsystems in accordance with the RT and the HRT conjectures.
It was observed in \cite {Hartman:2013qma} that the eternal black hole in $AdS_3$ may be mapped to a BTZ black brane geometry. In this case it was demonstrated that 
for early times the extremal surface (geodesic) passes through the interior region of the black brane geometry connecting one asymptotic region to the other. Hence the linear growth of the entanglement entropy was related to the growth of the extremal surface along the {\it nice slice}\footnote {Nice slices are spacelike surfaces with small curvatures where any matter moves with modest velocity in the local frame defined by these slices \cite{Polchinski:1995ta}.} in the interior region of the black brane. For late times the extremal surface reduces to a static minimal surface located exterior to the horizon and the entanglement entropy is rendered extensive \cite{Hartman:2013qma}. It is argued in \cite{Hartman:2013qma} that the holographic entanglement entropy of the configuration in question after a global quench scenario \cite{Calabrese:2005in} is just half of the corresponding holographic entanglement entropy for the eternal black hole geometry.

Quite naturally the developments described above lead to the interesting issue of a holographic characterization for the time evolution of the entanglement negativity for mixed state configurations following a global quench in a dual $CFT_{1+1}$.  As mentioned earlier the correct holographic dual for the global quench scenario is described by an eternal black hole geometry sliced in half by an ETW brane. Hence to this end it is required to obtain the holographic entanglement negativity for such mixed states from the dual eternal black hole geometry. 

In this article we address the significant issue described above and investigate the time evolution of the holographic entanglement negativity for mixed state configurations of disjoint and adjacent intervals following a global quench in a $CFT_{1+1}$. As described in \cite{Coser:2014gsa} the entanglement negativity for the above configurations in a $CFT_{1+1}$ with a global quench involve the corresponding four and three point twist correlators on a strip. It is significant to mention here that the universal parts of these twist correlators relevant in the space time scaling limit factorize into products of various two point twist correlators on a strip. This leads to an expression for the entanglement negativity of these configurations which involve a specific algebraic sum of the corresponding entanglement entropies of appropriate intervals and their combinations. The holographic entanglement negativity for the above configurations may then be computed from the prescription for the entanglement entropies described in \cite{Hartman:2013qma} for a bulk eternal black hole geometry sliced in half by an ETW brane. 
From our computations we observe an interesting structure of bulk geodesics connecting the two asymptotic regions which evolves with time to a geodesic structure that is located only in the exterior region. The corresponding holographic entanglement negativity for the mixed state configurations following a global quench in a $CFT_{1+1}$ is then given by half of that obtained for the eternal black hole geometry. It is observed that the holographic entanglement negativity for the mixed state configurations in question exhibits a linear growth during the initial part of its time evolution. For late times it decreases linearly with an identical slope from a maximum value and goes to zero after a certain value of time which indicates the process of thermalization and black hole formation in the bulk \cite{Calabrese:2006rx,Balasubramanian:2010ce,Balasubramanian:2011ur,Buchel:2013lla}. Remarkably the holographic entanglement negativity of the mixed state configurations in question, computed through our construction exactly reproduces the corresponding $CFT_{1+1}$ replica technique results following a global quench as described in \cite{Coser:2014gsa}.

This article is organized as follows. In Sec. (\ref{eeandn}), we briefly review the time evolution of entanglement entropy for a single interval after a global quench in a $CFT_{1+1}$. The evolution of entanglement negativity for two disjoint and adjacent intervals after a global quench in a $CFT_{1+1}$ is reviewed in Sec. (\ref{enandn}). In Sec. (\ref{heeag}) we review the time evolution of the corresponding holographic entanglement entropy for a single interval from a dual bulk eternal black hole geometry. In the next Sec. (\ref{henaq}), we establish our holographic construction for the 
time evolution of the entanglement negativity for two disjoint and adjacent intervals from the dual bulk eternal black hole geometry. The holographic entanglement negativity for the mixed state configurations following a global quench is then given by half of that obtained for the eternal black hole geometry and exactly matches with the corresponding $CFT_{1+1}$ replica technique results. We then present a summary and our conclusions in the final Sec. (\ref{sumanddis}).

\section{Entanglement entropy}\label{eeandn}
In this section we briefly review the time evolution of the entanglement entropy following a global quench as described in \cite{Calabrese:2005in,Calabrese:2009qy,calabrese2016quantum,Coser:2014gsa}. In a global quench scenario, the system is initially prepared in a pure state $| \psi_0 \rangle$ which is the ground state of a Hamiltonian $H_0$ at time $t=0$. For time $t>0$ the same system is allowed to evolve unitarily with respect to another Hamiltonian $H$, where $H$ and $H_0$ are related to each other by an experimentally tunable parameter.  The unitary evolution of the density matrix $\rho_0=|\psi_0 \rangle  \langle \psi_0 |$ for $t>0$ is governed by the post-quenched Hamiltonian $H$ as
\begin{equation}
\rho(t) = | \psi(t) \rangle  \langle \psi(t) |=e^{-iHt}| \psi_0 \rangle  \langle \psi_0 |e^{iHt},
\end{equation}
where $\rho(t)$ is the density matrix at a time $t$ which is always in a pure state as a consequence of unitary evolution. The entanglement entropy of a subsystem $A$ is obtained as the von Neumann entropy of the reduced density matrix 
$\rho_{A}(t)={\rm Tr}_B \rho(t)$ where $B=A^c$ represents the rest of the system, which is given as
\begin{equation}\label{ent_entropy}
 S_A(t)=- \mathrm{Tr}\left(\rho_{A}(t)\ln \rho_{A}(t)\right).
\end{equation}

\begin{figure}[H]
\centering
\includegraphics[scale=.3]{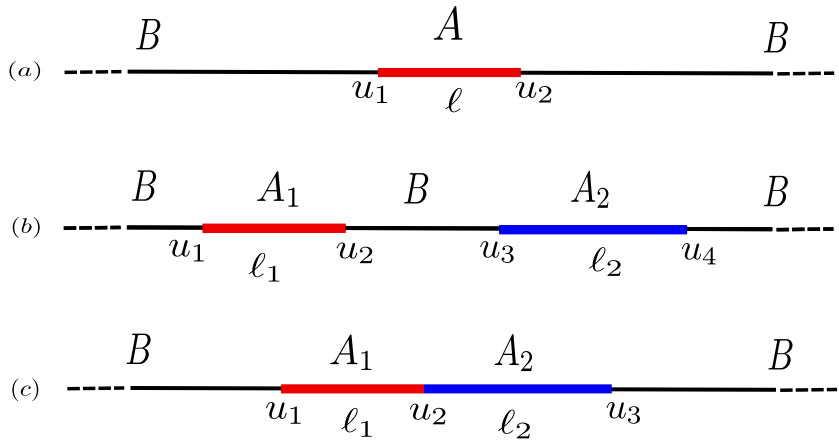}
\caption{Configuration of a (a) single interval (top), (b) two disjoint intervals (middle), and (c) two adjacent intervals (bottom) in a $(1+1)$-dimensional ${CFT}$.}\label{3intervals}
\end{figure}

In a $CFT_{1+1}$  for simplicity the initial state may be chosen as  $| \psi_0 \rangle=e^{-{\tau_0 H}/{4}} |B \rangle$ where $|B\rangle$ is a conformal invariant boundary state (B-state) and $\tau_0$ is the correlation length in the initial state. The entanglement entropy $S_A(t)$ may be obtained through a replica technique as follows
\begin{equation}\label{eeoneint}
 S_A(t)=-\lim_{n\to 1}\frac{\partial}{\partial n}{\rm Tr}  \rho_A^n.
\end{equation} 
The quantity ${\rm Tr}  \rho_A^n$ for a single interval $A$ of length $\ell$ as depicted in Fig. \ref{3intervals}(a) in the $CFT_{1+1}$, is related to the two point twist correlator on a strip of width $2\tau_0$ as \cite{Calabrese:2005in,Calabrese:2009qy,calabrese2016quantum,Coser:2014gsa}
 \begin{equation}
 \mathrm{Tr} \rho_A^n =\langle\,\mathcal{T}_n (w_{1}) \bar{\mathcal{T}}_n(w_{2}) \,\rangle_{\rm strip} \,,
\end{equation}
 where $w_i= u_i + \textrm{i} \tau \,$ are the complex coordinates on the strip ($u_i \in \mathbb{R}$ and $0< \tau< 2\tau_0$). The above twist correlator may be obtained through a conformal map $z_i=\exp(\pi w_i/2\tau_0)$ from the strip to the upper half plane (UHP) which is then given as 
\begin{equation}\label{renyi corr strip}
 \langle\,\mathcal{T}_n (w_{1}) \bar{\mathcal{T}}_n(w_{2}) \,\rangle_{\rm strip} \,=\,
c_n \left( \frac{\pi}{2\tau_0} \right)^{2\Delta_n}{\frac{1}{|(z_1 - \bar{z}_1)(z_2 - \bar{z}_2) \,\eta_{1,2}|^{\Delta^{}_n}}} {\cal F}(\eta_{1,2})\,,
 \end{equation}
where $c_n$ is a constant and $\eta_{1,2} $ is the cross ratio, and the scaling dimension $\Delta_n$ of the twist operators $\mathcal{T}_n$ and $\bar{\mathcal{T}}_n$ is given as follows
 \begin{equation}
 \label{Delta_n def}
\Delta_n =   \frac{c}{12}\left( n- \frac{1}{n} \right) .
 \end{equation}
The function ${\cal F}_{n}(\eta_{1,2})$ in eq. \eqref{renyi corr strip} is non-universal and depends on the full operator content of the theory. Note that only the limiting behavior of $ \eta_{1,2}\to 0 $ and $ \eta_{1,2}\to 1 $ are important and in these limits the non-universal function ${\cal F}_{n}(\{\eta_{1,2}\}$ is just a constant \cite{Coser:2014gsa}. The entanglement entropy $S_A$ may now be obtained by utilizing eqs. (\ref{eeoneint}) and \eqref{renyi corr strip} and considering the {\em spacetime scaling limit} ($t, \ell \gg \tau_0$), which is given as follows (see \cite{Coser:2014gsa} for detailed computation)
\begin{equation}
\label{SA one interval}
S_A = \frac{\pi c}{6\tau_0}\big[t + q(t, \ell) \big]= 
\left\{\begin{array}{ll}\displaystyle 
\frac{\pi c}{6\tau_0}t \hspace{2cm}& t <\ell/2\,, \\ \\ \displaystyle 
\frac{\pi c}{12\tau_0} \ell & t >\ell/2\,,
\end{array}\right.
\end{equation}
where the function $q(t, \ell)$ is defined as follows
\begin{equation}\label{qfns}
q(t,\ell) \,\equiv\, \frac{\ell}{2} -  \textrm{max}(t, \ell/2)
\,=\,
\left\{\begin{array}{ll}
0 \hspace{2cm}& t <\ell/2\,, \\
\ell/2-t  & t >\ell/2\,.
\end{array}\right.
\end{equation}
Note that the entanglement entropy $S_A$ increases linearly with time for $t<\ell/2$ and saturates to a thermal value beyond the time $t\sim\ell/2$. It may be observed that the entanglement entropy for large times is same as the thermal entropy of a mixed state at a large but finite temperature $T=\frac{1}{\beta}=\frac{1}{4\tau_0}$. This saturation to a thermal entropy at large times leads to the fascinating phenomena of $CFT$ thermalization as described in\cite{Calabrese:2006rx,Coser:2014gsa}.

\section{Entanglement negativity}\label{enandn}
In this section we briefly recapitulate the evolution of the entanglement negativity for mixed states of two disjoint and adjacent intervals in a $CFT_{1+1}$ after a global quench scenario as described in \cite{Coser:2014gsa}. As mentioned in the introduction, for a bipartite system $A=A_1\cup A_2$ in a mixed state, the entanglement between the subsystems $A_1$ and $A_2$ is characterized by the entanglement negativity, which is defined as \cite{PhysRevA.65.032314}
\begin{equation}\label{ent_neg}
\mathcal{E} = \ln \mathrm{Tr}|\rho_A^{T_2}|.
\end{equation}
Here the partial transposed reduced density matrix $\rho_A^{T_2}$ is obtained as follows
\begin{equation}\label{26}
\langle e^{(1)}_ie^{(2)}_j|\rho_A^{T_2}|e^{(1)}_ke^{(2)}_l\rangle = 
\langle e^{(1)}_ie^{(2)}_l|\rho_A|e^{(1)}_ke^{(2)}_j\rangle,
\end{equation}
where $|e^{(1)}_i\rangle$ and $|e^{(2)}_j\rangle$ are the bases for the the Hilbert spaces $\mathcal{H}_{A_1}$ and  $\mathcal{H}_{A_2}$ of the subsystems $A_1$ and $A_2$ respectively. In a $CFT_{1+1}$ the entanglement negativity is obtained through a suitable replica technique through the quantity 
$\mathrm{Tr} \big(\,\rho_A^{T_{2}}\big)^{n}$ for even $n=n_e$ and the analytic continuation
to $n_e\to 1$ as follows \cite{Calabrese:2012ew,Calabrese:2012nk,Calabrese:2014yza}
\begin{equation}\label{en replica}
\mathcal{E} = \lim_{n_e \rightarrow 1}  \ln \Big[ \mathrm{Tr} \big(\,\rho_A^{T_{2}}\big)^{n_e} \Big]\,.
\end{equation}

\subsection{Two disjoint intervals}\label{ent_neg_glo_quench_dj_int}
Now consider a bipartite system $A=A_1 \cup A_2$ of two disjoint intervals $A_1$ and $A_2$ in a $CFT_{1+1}$ of lengths $\ell_1=|u_2-u_1|$ and $\ell_2=|u_4-u_3|$ separated by a distance $\ell_s=|u_3-u_2|$ as depicted in Fig. \ref{3intervals}(b). In this case the quantity  $\mathrm{Tr} (\rho_A^{T_2})^{n}$ is given by a four-point twist correlator on a strip of width $2\tau_0$ as  \cite{Coser:2014gsa}
\begin{equation}\label{4pttc}
\begin{aligned}
\mathrm{Tr}(\rho_A^{T_2})^{n} = &
\langle\mathcal{T}_{n}(w_1)\overline{\mathcal{T}}_{n}(w_2)\overline{\mathcal{T}}_{n}(w_3)\mathcal{T}_{n}(w_4)\rangle_{\rm strip}\\
=&\,
c_n^2
\bigg(\frac{\pi}{2\tau_0} \bigg)^{\Delta}
 \prod_{i=1}^{4} \bigg| \frac{z_i}{z_i - \bar{z}_i} \bigg|^{\Delta_n}
\frac{1}{\eta_{1,2}^{\Delta_n}\,\eta_{3,4}^{\Delta_n}}
\left(
\frac{\eta_{1,4}\,\eta_{2,3}}{\eta_{1,3}\,\eta_{2,4}}
\right)^{\Delta^{(2)}_n/2-\Delta_n} {\cal F}(\{\eta_{j,k}\}) \, ,
\end{aligned}
\end{equation}
where $\Delta = 4\Delta_n$ and $z_i = e^{\pi (w_i / (2\tau_0)}$ with $w_i=u_i+i\tau$ are the complex coordinates on the strip. The scaling dimension $\Delta^{(2)}_n $ of the twist operators $\mathcal{T}^2_n$ and $\bar{\mathcal{T}}_n^2$ in eq. \eqref{4pttc}  is given as 
\begin{equation}
\label{delta2 def}
\Delta^{(2)}_n \equiv  
\left\{ \begin{array}{ll}
\displaystyle
\Delta_{n}
\hspace{.5cm}& 
\textrm{odd $n$\,,}
\\
\rule{0pt}{.5cm}
\displaystyle
2\Delta_{n/2}
\hspace{.5cm}& 
\textrm{even $n$\,,}
\end{array}
\right.
\end{equation}
where $\Delta_{n}$ is defined in (\ref{Delta_n def}).

The time evolution of the entanglement negativity $\cal{E}$ may then be obtained by utilizing the eqs. (\ref{en replica}) and (\ref{4pttc}) considering the {\em spacetime scaling limit} ($t, \ell \gg \tau_0$) as   \cite{Coser:2014gsa}
\begin{equation}\label{logneg N2disj cft t-dep}
\begin{aligned}
\mathcal{E} = &\frac{\pi c}{8\tau_0}\left[q(t,u_3-u_1) + q(t,u_4-u_2) - q(t,u_4-u_1) - q(t,u_3-u_2) \right],\\
=&\frac{\pi c}{8\tau_0}\left[q(t,\ell_1+\ell_s) + q(t,\ell_2+\ell_s) - q(t,\ell_1+\ell_2+\ell_s) - q(t,\ell_s) \right],
\end{aligned}
\end{equation}
where the function $q(t,u_i-u_j)$ with $(i,j \in 1, 2, 3, 4)$ is defined in the eq. (\ref{qfns}). Note that the function $q(t,u_i-u_j)$ depends non trivially on the relative values of the time $t$ and the lengths of the intervals $\ell_1$, $\ell_2$ and $\ell_s$ leading to various interesting limits for the entanglement negativity in this case. We have computed the entanglement negativity in this case for each of these significant limits as which traces the time evolution of this quantity as described below.

\subsubsection{$t<\ell_s/2<\ell_1/2<\ell_2/2$}
In this limit of early time none of the functions $q(t,u_i-u_j)$s defined in eq. (\ref{qfns}) contribute in eq. (\ref{logneg N2disj cft t-dep}) and consequently the entanglement negativity vanishes
\begin{equation}\label{early_time_evo_dis_joint}
\mathcal{E}\,=\,0.
\end{equation}

\subsubsection{$\ell_1/2<\ell_2/2<t<(\ell_1+\ell_s)/2$}
For this limit only the function $q(t,u_3-u_2)$ contributes and the other $q(t,u_i-u_j)$s vanish in eq. (\ref{logneg N2disj cft t-dep}), leading to a non-zero time dependent value of the entanglement negativity which is given as follows
\begin{equation}\label{initial_time_evo_dis_joint}
\mathcal{E}\,=\,\frac{\pi c}{8\tau_0}\left(t-\frac{\ell_s}{2} \right)\,.
\end{equation}

\subsubsection{$\ell_2/2<(\ell_1+\ell_s)/2<t<(\ell_2+\ell_s)/2$}
In this case the functions $q(t,u_3-u_1)$ and $q(t,u_3-u_2)$ are non-zero in eq. (\ref{logneg N2disj cft t-dep}), and the entanglement negativity saturates to a constant value proportional to the length of the interval $A_1$ given by
\begin{equation}\label{intermediate_time_evo_dis_joint}
\mathcal{E}\,=\,\frac{\pi c\ell_1}{16\tau_0}\,.
\end{equation}

\subsubsection{$(\ell_1+\ell_s)/2<(\ell_2+\ell_s)/2<t<(\ell_1+\ell_2+\ell_s)/2$}

In this limit the functions $q(t,u_3-u_1)$, $q(t,u_4-u_2)$ and $q(t,u_3-u_2)$ are non-zero in eq. (\ref{logneg N2disj cft t-dep}), and the entanglement negativity may be expressed as
\begin{equation}\label{large_time_evo_dis_joint}
\mathcal{E}\,=\,\frac{\pi c}{8\tau_0}\left(\frac{\ell_1+\ell_2+\ell_s}{2}-t \right)\,.
\end{equation}

\subsubsection{$(\ell_1+\ell_s)/2<(\ell_2+\ell_s)/2<(\ell_1+\ell_2+\ell_s)/2<t$}
For this limit of late times all of the functions $q(t,u_i-u_j)$s in the eq. (\ref{logneg N2disj cft t-dep}) contribute but the entanglement negativity vanishes due to mutual cancellations
\begin{equation}\label{final_time_evo_dis_joint}
\mathcal{E}\,=\,0.
\end{equation}

The above results, obtained by us for the time evolution of the entanglement negativity following a global quench, may be described as follows. The entanglement negativity for early time described in eq. (\ref{early_time_evo_dis_joint}) is zero and increases linearly during the initial period of the time evolution as described in eq. (\ref{initial_time_evo_dis_joint}). This saturates to a constant value described in eq. (\ref{intermediate_time_evo_dis_joint}) for an intermediate range of time. As time increases further the entanglement negativity decreases linearly as described in eq. (\ref{large_time_evo_dis_joint}) and vanishes at late times describing the thermalization of the mixed state under consideration in the $CFT_{1+1}$.

\subsection{Two adjacent intervals}\label{ent_neg_glo_quench_adj_int}

Having described the case for the mixed state of disjoint intervals above we now proceed
to analyze the case for the mixed state of adjacent intervals $A_1$ and $A_2$. For this case the quantity  
$\mathrm{Tr} (\rho_A^{T_2})^n$ is given by a three-point twist correlator on a strip as follows  \cite{Coser:2014gsa}
\begin{equation}\label{en 3pt fn}
\begin{aligned}
\mathrm{Tr} (\rho_A^{T_2})^n
 =&\langle \mathcal{T}_n(w_1) \bar{\mathcal{T}}^2_n(w_2) \mathcal{T}_n(w_3) \rangle_{\rm strip}\,\\
 =&\,
 c_n
\bigg(\frac{\pi}{2\tau_0} \bigg)^{\Delta}
  \prod_{i=1}^{3} \bigg| \frac{z_i}{z_i - \bar{z}_i} \bigg|^{\Delta_{(i)}}
\left(
\frac{\eta_{1,3}^{\Delta_n^{(2)}-2\Delta_n}}{\eta_{1,2}^{\Delta_n^{(2)}}  \eta_{2,3}^{\Delta_n^{(2)}}}
\right)^{1/2} {\cal F}(\{\eta_{j,k}\}) \, ,
\end{aligned}
\end{equation}
where $\Delta = 2\Delta_n+\Delta^{(2)}_n$. The entanglement negativity ${\cal E}$ may then be obtained from eqs. (\ref{en 3pt fn}) and (\ref{en replica}) by employing the {\em spacetime scaling limit} ($t, \ell \gg \tau_0$) as  \cite{Coser:2014gsa} 
\begin{equation}\label{en adj cft}
\begin{aligned}
\mathcal{E}\,=&\,\frac{\pi c}{8\tau_0}\,\big[\, t  - q(t,u_3-u_1) + q(t,u_2-u_1) + q(t,u_3-u_2)  \big]\,,\\
=&\,\frac{\pi c}{8\tau_0}\,\big[\, t  - q(t,\ell_1+\ell_2) + q(t,\ell_1) + q(t,\ell_2)  \big]\,.
\end{aligned}
\end{equation}
In the above expression as earlier the function $q(u_i, u_j)$ in eq. (\ref{qfns}) non trivially depends on the relative values of the time $t$ and the lengths of the intervals $\ell_1$ and $\ell_2$ leading to certain significant limits. In what follows, as earlier we compute the entanglement negativity of the mixed state in question for these interesting limits which describes the time evolution of this measure.

\subsubsection{$t<\ell_1/2<\ell_2/2<(\ell_1+\ell_2)/2$}
For this early time the functions $q(t,u_i-u_j)$s in the eq. (\ref{en adj cft}) vanish and the entanglement negativity may then be given as
\begin{equation}\label{early_time_evo_adj_joint}
\mathcal{E}\,=\,\frac{\pi c }{8\tau_0}t\,.
\end{equation}
\subsubsection{$\ell_1/2<t<\ell_2/2<(\ell_1+\ell_2)/2$}
In this case only the function $q(t,u_2-u_1)$ is non-zero in the eq. (\ref{en adj cft}) and the entanglement negativity is proportional to the length of the subsystem $A_1$ as
\begin{equation}\label{initial_time_evo_adj_joint}
\mathcal{E}\,=\,\frac{\pi c }{16\tau_0}\ell_1\,.
\end{equation}

\subsubsection{$\ell_1/2<\ell_2/2<t<(\ell_1+\ell_2)/2$}
In this case the functions $q(t,u_2-u_1)$ and $q(t,u_3-u_2) $ in the eq. (\ref{en adj cft}) are non-zero leading to a non-zero time dependent value of the entanglement negativity given as follows
\begin{equation}\label{late_time_evo_adj_joint}
\mathcal{E}\,=\,\frac{\pi c }{8\tau_0}\left(\frac{\ell_1+\ell_2}{2}-t\right)\,.
\end{equation}
\subsubsection{$\ell_1/2<\ell_2/2<(\ell_1+\ell_2)/2<t$}
In the limit of late time all of the functions $q(t,u_i-u_j)$s in the eq. (\ref{en adj cft}) are non-zero which leads to the vanishing value of the entanglement negativity 
\begin{equation}\label{final_time_evo_adj_joint}
\mathcal{E}\,=\,0\,.
\end{equation}

The evolution of the entanglement negativity from the above results is described  as follows. It is observed that at
the initial part of of its time evolution the entanglement negativity increases linearly with time described in eq. (\ref{early_time_evo_adj_joint}) and saturates to a constant value for an intermediate time as given in eq. (\ref{initial_time_evo_adj_joint}). As time progress further negativity decreases linearly as described in eq. (\ref{late_time_evo_adj_joint}) and vanishes at late times indicating the phenomena of thermalization of the mixed state in consideration.
\section{Holographic entanglement entropy}\label{heeag}
In this section we briefly review the holographic computation of the entanglement entropy for a single interval in a $CFT_{1+1}$ following a global quench in the $AdS_3/CFT_2$ framework. As described in \cite{Hartman:2013qma} the correct holographic dual of the global quench scenario is given by the single sided eternal black hole obtained by slicing the Penrose diagram of the usual eternal black hole in half with an end of the world (ETW) brane. In this instance the minimal area extremal surfaces (geodesics for $AdS_3/CFT_2$) terminate on the ETW brane and the corresponding holographic entanglement entropy is just half of that obtained for the usual eternal black hole geometry. 

\begin{figure}[h]
\centering
\includegraphics[width=.75\linewidth]{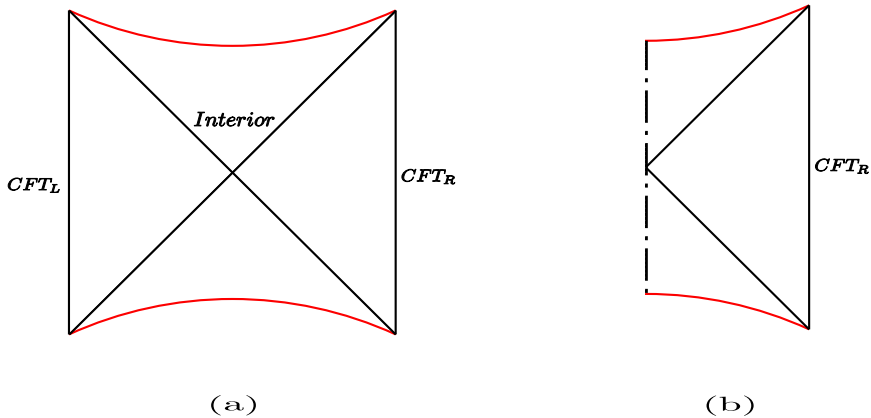}
\caption{$(a)$ Penrose diagram of the eternal black hole and $(b)$ the eternal black hole cut in half by an end of the world (ETW) brane (Figure modified from that in \cite{Hartman:2013qma}).}\label{Penrosediagrameternal}
\end{figure}

\begin{figure}[h]
\centering
\includegraphics[width=.75\linewidth]{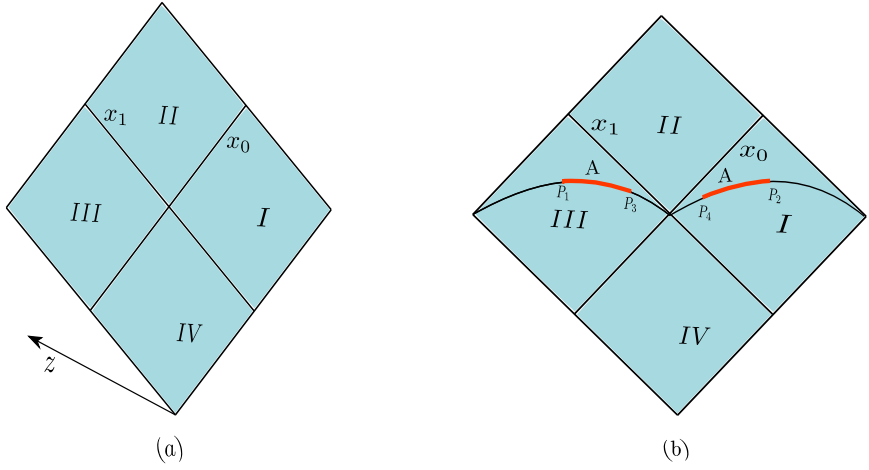}
\caption{$(a)$ Different regions of the BTZ black string. Region $I$ and $III$ are Rindler wedges. $(b)$ Endpoints of the interval $A$ on both sides of the Rindler wedges (Figure modified from that in \cite{Hartman:2013qma}).}\label{patch}
\end{figure}

In the above context we begin by reviewing the time evolution of the holographic entanglement entropy from a bulk eternal BTZ black hole geometry described by a two-sided Penrose diagram as depicted in Fig. \ref{Penrosediagrameternal}. As described in \cite{Hartman:2013qma,Maldacena:1998bw} this bulk geometry is dual to the two copies of the $CFT$ ($CFT_R\otimes CFT_L$), in the thermofield double state\footnote {Thermofield double state is a purification of a mixed state at some temperature. This is a particular pure but non-trivially entangled state in the full Hilbert space of the CFT. }. To obtain the holographic entanglement entropy of a single interval $A$ in this scenario, it is required to consider two copies of $A$ in the two $CFT$s defined on either side of the Penrose diagram as shown in Fig. \ref{patch}(b). It is well known that the bulk eternal BTZ black hole is a quotient space of $AdS_3$ and hence the angular coordinate may be unwrapped to map the  
the two boundaries of the BTZ Penrose diagram to the corresponding Rindler wedges \cite{Hartman:2013qma,Maldacena:1998bw} as described in Fig. \ref{patch}(a). The bulk geometry in this case is then an eternal BTZ black string.
It is now useful to describe the different regions of the BTZ black string and relate each coordinate patch to the Poincar\'e coordinates,
\begin{equation}\label{Poincare_coordinate}
ds^2 = \frac{1}{z^2}(-dx_0^2 + dx_1^2 + dz^2).
\end{equation}
The exterior metric of the BTZ black string is given as
\begin{equation}\label{BTZ_exterior}
ds^2 = -\sinh^2\rho dt^2 + \cosh^2\rho dx^2 + d\rho^2,
\end{equation}
where the horizon is located at $\rho=0$, and the boundary is at $\rho\rightarrow\infty$. The BTZ coordinates $(t,x,\rho)$ covers a part of the Poincar\'e patch, close to the boundary and are related to the Poincar\'e coordinates as follows
\begin{equation}\label{coordinterval}
x_1\pm x_0 \approx   e^{-x \pm t} \ , \quad\quad  {1\over z} \approx {1\over 2} e^{\rho - x}.
\end{equation}
The coordinates $(t,x)$ cover the Rindler wedge denoted as the region $I$ of the Minkowski diamond in Fig. \ref{patch}(a). It is possible to reach the other Rindler wedge denoted by the region $III$ in Fig. \ref{patch}(a) by analytically continuing the time $t \to t + i \pi$. This is the continuation through which it is possible to reach from one side of the exterior region of the black string Penrose diagram to the other.
The interior metric to the horizon $\rho=0$ is obtained by analytically continuing the coordinates $\rho = i\alpha, t = \tilde{t} - i\pi/2$ in eq. (\ref{BTZ_exterior}) as follows
\begin{equation}\label{BTZ_interior}
ds^2 = \sin^2\alpha d\tilde t^2 + \cos^2\alpha dx^2 - d\alpha^2.
\end{equation}
These coordinates cover some part of the interior region and meet the boundary along the light cone $x_0^2-x_1^2=0$ as depicted in Fig. \ref{patch}(a). The corresponding future bulk region of the interior to the BTZ black string is obtained by setting $\alpha = \pi/2 - i \tilde{\rho}, x = \tilde{x} - i \pi/2$ in eq. (\ref{BTZ_interior}), which is given as
\begin{equation}\label{BTZ_future}
ds^2 = \cosh^2\tilde{\rho}d\tilde t^2 - \sinh^2\tilde\rho d\tilde x^2 + d\tilde\rho^2 .
\end{equation}
The boundary of this future region in the Poincar\'e coordinates is given as $z\leq x_0^2-x_1^2$.

\begin{figure}
\centering
\includegraphics[width=.7\linewidth]{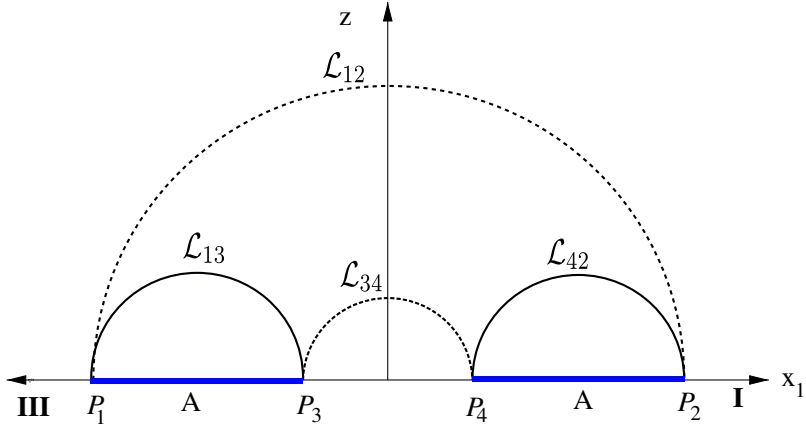}
\caption{Geodesics configuration for a single interval projected into the $x_1$, $z$ plane. For early time geodesics are shown by the dotted lines and for large time geodesics are shown by the solid lines (Figure modified from that in \cite{Hartman:2013qma}).}\label{eegeo}
\end{figure}

We now consider the interval $A$ of length $\ell$ on both the Rindler wedges $I$ and $III$
of the Minkowski diamond in Fig. \ref{patch}(b). These two intervals are described as 
$(P_1,P_3)$ and $(P_2,P_4)$ in the $x_1$-$z$ plane of the Poincar\'e coordinates as depicted in Fig. (\ref{eegeo}). The coordinates of the end points of the intervals at a time $t$ are given from eq. \eqref{coordinterval} as follows \cite{Hartman:2013qma}
\begin{equation}
\begin{aligned}
P_1=&(\sinh t, -\cosh t),~~~P_3=e^{-\ell} P_1,\\
P_2=&(\sinh t, \cosh t),~~~~~~P_4=e^{-\ell} P_2.
\end{aligned}
\end{equation}
At early time $t\leq \ell/2$, the entanglement entropy receives contribution form the lengths of the geodesics $\mathcal{L}_{12}$ and $\mathcal{L}_{34}$ between the points $P_1$ to $P_3$ and $P_2$ to $P_4$ respectively as depicted in Fig. \ref{eegeo}. These geodesics pass through the interior region in eq. (\ref{BTZ_interior}) of the BTZ black string connecting one Rindler wedge to the other (the dashed  geodesics in Fig. \ref{eegeo}). These two geodesics are semicircles and they have identical regularized lengths. The entanglement entropy at early time $t\leq \ell/2$ is then given as follows  \cite{Hartman:2013qma}
\begin{equation}\label{earlytimes}
\begin{aligned}
S_A^{(1)}=&\frac{1}{4G_N}(\mathcal{L}_{12}+\mathcal{L}_{34}),\\
=&\frac{4\pi c t}{3\beta}+4S_{div}.
\end{aligned}
\end{equation}
Here $S_{div}=\frac{c}{6}\ln\left(\frac{\beta}{4\pi\epsilon}\right)$ is the divergent part of the entanglement entropy, $\beta=4\tau_0$ is the inverse temperature and $\epsilon$ is the UV cut off. 
On the other hand for times $t>\ell/2$ the entanglement entropy receives contribution   from the lengths of the geodesics $\mathcal{L}_{13}$ and $\mathcal{L}_{42}$ between the points $P_1$ to $P_3$ and $P_2$ to $P_4$ respectively. These geodesics are confined to the exterior region of the BTZ black string outside the horizon as described in eq. (\ref{BTZ_exterior}) (depicted by the solid lines in the Fig. \ref{eegeo}). The entanglement entropy for time $t>\ell/2$ is then given as
\begin{equation}\label{latetimes}
\begin{aligned}
S_A^{(2)}=&\frac{1}{4G_N}(\mathcal{L}_{13}+\mathcal{L}_{24}),\\
=&\frac{2\pi c \ell}{3\beta}+4S_{div}.
\end{aligned}
\end{equation}

It is observed from the above development that at early time the geodesics pass through the interior region of the BTZ black brane geometry from one Rindler wedge $I$ to the other $III$ and the linear increase of the holographic entanglement entropy is related to the growth of the {\it nice slice} in the interior. On the other hand for late times the geodesics are confined to the exterior region sitting outside the horizon and the corresponding entanglement entropy becomes thermal.

As mentioned earlier the bulk dual geometry corresponding to a global quench in a $CFT_{1+1}$ where the initial state is the conformal invariant boundary state $|B\rangle$ \cite {Calabrese:2005in} is described by the eternal black hole sliced in half by an ETW brane (single sided black hole). Hence the corresponding holographic entanglement entropy for
the interval $A$ in this case is then described by half of that obtained in eqs. (\ref{earlytimes}) and (\ref{latetimes}) respectively. Remarkably, these results match exactly with the corresponding $CFT_{1+1}$ computation through the replica technique in eq. (\ref{SA one interval}).
\section{Holographic entanglement negativity}\label{henaq}
In this section we present our holographic construction for the time evolution of the entanglement negativity of two disjoint and adjacent intervals in a $CFT_{1+1}$ following a global quench as described in \cite{Coser:2014gsa}. Motivated by the holographic construction for the time evolution of the entanglement entropy in \cite{Hartman:2013qma} reviewed above, we begin by computing the holographic entanglement negativity for the mixed states under consideration from the bulk dual eternal black hole geometry. The corresponding holographic entanglement negativity following a global quench may then be obtained by considering half of the result for the bulk dual eternal black hole geometry. 

It interesting to note that the four point twist correlator in eq. \eqref{4pttc} admits a factorization with the non universal function ${\cal F}(\{\eta_{j,k}\})\to 1$ for the cross ratios $\eta_{i,j}\to 1$ and $\eta_{i,j}\to 0$, as follows
\begin{equation}\label{factorization_four_pt}
\begin{aligned}
&\langle\mathcal{T}_{n}(w_1)\overline{\mathcal{T}}_{n}(w_2)\overline{\mathcal{T}}_{n}(w_3)\mathcal{T}_{n}(w_4)\rangle_{\rm strip}=\\
&\frac{\langle\mathcal{T}_{n}(w_1)\overline{\mathcal{T}}_{n}(w_4)\rangle_{\rm strip}~\langle\mathcal{T}_{n}(w_2)\overline{\mathcal{T}}_{n}(w_3)\rangle_{\rm strip}~\Big(\langle\, \mathcal{T}^2_n (w_1) 
\bar{\mathcal{T}}^2_n(w_3) \,\rangle_{\rm strip}~\langle\, \mathcal{T}^2_n (w_2) 
\bar{\mathcal{T}}^2_n(w_4) \,\rangle_{\rm strip}\Big)^\frac{1}{2}}{\langle\mathcal{T}_{n}(w_1)\overline{\mathcal{T}}_{n}(w_3)\rangle_{\rm strip}~\langle\mathcal{T}_{n}(w_2)\overline{\mathcal{T}}_{n}(w_4)\rangle_{\rm strip}~\Big(\langle\, \mathcal{T}^2_n (w_1) 
\bar{\mathcal{T}}^2_n(w_4) \,\rangle_{\rm strip}~\langle\, \mathcal{T}^2_n (w_2) 
\bar{\mathcal{T}}^2_n(w_3) \,\rangle_{\rm strip} \Big)^\frac{1}{2}},
\end{aligned}
\end{equation}
where we have employed eq. \eqref{renyi corr strip} and the following two point twist correlator 
\begin{equation}
\label{twist squared uhp N=1}
\langle  \mathcal{T}^2_n (w_1) 
\bar{\mathcal{T}}^2_n(w_2) \rangle_{\rm strip} 
=\left( \frac{\pi}{2\tau_0} \right)^{2\Delta^{(2)}_n}
\frac{c^{(2)}_{n}}{|(z_1 - \bar{z}_1)(z_2 - \bar{z}_2) \,\eta_{1,2}|^{\Delta^{(2)}_n}} {\cal F}(\eta_{1,2})\,.
\end{equation}
The entanglement negativity for two disjoint intervals may then be obtained by utilizing eqs. \eqref{en replica} and \eqref{factorization_four_pt} which leads to following expression
\begin{equation}
\begin{aligned}
\label{eq_holo_ent_neg_dj_int_holo_ent_enpy}
{\cal E} = &\frac{ 3 }{ 4 }
\left ( S_{ A_1 \cup A_s } + S_{ A_s \cup A_2 }
- S_{ A_1 \cup A_2 \cup A_s } - S_{ A_s } \right ),
\end{aligned}
\end{equation}
where $S_{\gamma}~(\gamma \in  A_1 \cup A_s,~A_s \cup A_2,~A_1 \cup A_2 \cup A_s,~ A_s)$ is the entanglement entropy of an interval $\gamma$  as given in eq. \eqref{SA one interval}.

Note that a similar factorization also holds for the three point twist correlator defined in eq. \eqref{en 3pt fn} which is given as

\begin{equation}\label{factorization_three_pt}
\begin{aligned}
\langle \mathcal{T}_n(w_1) \bar{\mathcal{T}}^2_n(w_2) \mathcal{T}_n(w_3) \rangle_{\rm strip}\,=
&\Bigg(\frac{\langle\, \mathcal{T}^2_n (w_1) 
\bar{\mathcal{T}}^2_n(w_2) \,\rangle_{\rm strip}~\langle\, \mathcal{T}^2_n (w_2) 
\bar{\mathcal{T}}^2_n(w_3) \,\rangle_{\rm strip} \langle\mathcal{T}_{n}(w_1)\overline{\mathcal{T}}_{n}(w_3)\rangle_{\rm strip}^2 }{\langle\, \mathcal{T}^2_n (w_1) 
\bar{\mathcal{T}}^2_n(w_3) \,\rangle_{\rm strip} }\Bigg)^{1/2}.
\end{aligned}
\end{equation}
The entanglement negativity for two adjacent intervals may then be obtained by utilizing eqs. \eqref{en replica} and \eqref{factorization_three_pt} as follows
\begin{equation}\label{heecon}
\mathcal{E} =  \frac{3}{4}(S_{A_1}+S_{A_2}-S_{A_1\cup A_2}).
\end{equation}

In what follows we utilize eqs. \eqref{eq_holo_ent_neg_dj_int_holo_ent_enpy} and \eqref{heecon} to establish the holographic entanglement negativity and its time evolution for two disjoint and adjacent intervals in a $CFT_{1+1}$ after a global quench employing the prescription for the holographic entanglement entropy described in \cite{Hartman:2013qma}.

\subsection{Two disjoint intervals}\label{hendjagq}
\begin{figure}[ht]
\centering
\includegraphics[width=\linewidth]{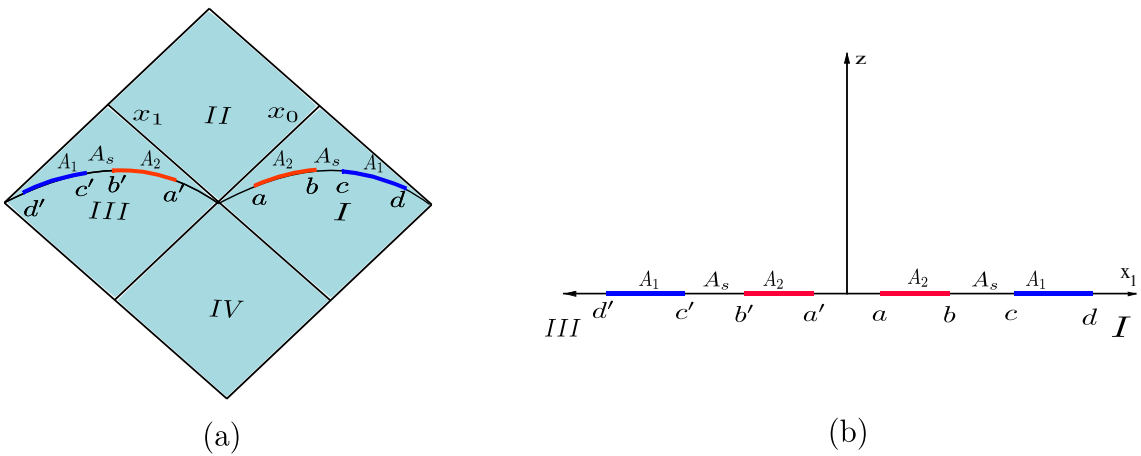}
\caption{(a) Two disjoint intervals $A_1$ (blue) and $A_2$ (red) on the Rindler wedges $I$ and $III$ respectively of the Minkowski diamond. (b) These two intervals are projected on the $x_1$-$z$ plane of the Poincar\'e coordinates.}\label{coord_disjoint_intervals}
\end{figure}

We first consider the case of two disjoint intervals $A_1$ and $A_2$ of lengths $\ell_1$ and $\ell_2$ with an interval $A_s$ of length $\ell_s$ separating them. As earlier for the dual eternal black hole geometry it is required to consider the intervals $A_1$, $A_2$ and $A_s$ in both the Rindler wedges $I$ and $III$ of the Minkowski diamond in Fig. \ref{coord_disjoint_intervals}(a). The end points of the intervals $A_1$, $A_2$ and $A_s$ in the Rindler wedges $I$ and $III$ projected on the $x_1$-$z$ plane of the Poincar\'e coordinates are denoted as ($a,b$), ($c,d$), ($b,c$) and ($a',b'$), ($c',d'$), ($b',c'$) respectively which is depicted in Fig. \ref{coord_disjoint_intervals}(b).

It is now possible to describe the time evolution of the holographic entanglement negativity for the mixed state configuration of the two disjoint intervals from the bulk dual eternal black hole geometry by utilizing  eqs. \eqref{eq_holo_ent_neg_dj_int_holo_ent_enpy}, \eqref{earlytimes} and \eqref{latetimes}. It is observed that distinct sets of geodesics contribute to the holographic entanglement negativity for various values of time $t$ relative to the lengths of the intervals $\ell_1$, $\ell_2$ and $\ell_s$ . In what follows we consider these significant limits and describe the relevant geodesic structures leading to the holographic entanglement negativity for each of these scenarios.

\subsubsection{$t<\ell_s/2<\ell_1/2<\ell_2/2$}

In the limit of early times all the relevant geodesics pass through the interior region of the BTZ black string connecting one Rindler wedge $I$ to the other $III$,  depicted by dashed curves in the Fig. \ref{neggeodjfin}(a). The corresponding holographic entanglement negativity is obtained by substituting the holographic entanglement entropy for the early time limit given in  eq. \eqref{earlytimes} in the eq. \eqref{eq_holo_ent_neg_dj_int_holo_ent_enpy}, which vanishes due to mutual cancellations as follows
\begin{equation}\label{djlima}
\begin{aligned}
\mathcal{E} = &\frac{3}{16G^{(3)}_N}\left(\underbrace{\mathcal{L}_{d'd}+\mathcal{L}_{b'b}}_{S_{ A_1 \cup A_s }}+\underbrace{\mathcal{L}_{c'c}+\mathcal{L}_{a'a}}_{S_{ A_s \cup A_2 }}   -\underbrace{\mathcal{L}_{d'd}-\mathcal{L}_{a'a}}_{S_{ A_1 \cup A_2 \cup A_s }}-  \underbrace{\mathcal{L}_{c'c}-\mathcal{L}_{b'b}}_{S_{ A_s }}\right),\\
=&0.
\end{aligned}
\end{equation}
\subsubsection{$\ell_1/2<\ell_2/2<t<(\ell_1+\ell_s)/2$}
\begin{figure}[h]
\centering
\includegraphics[width=\linewidth]{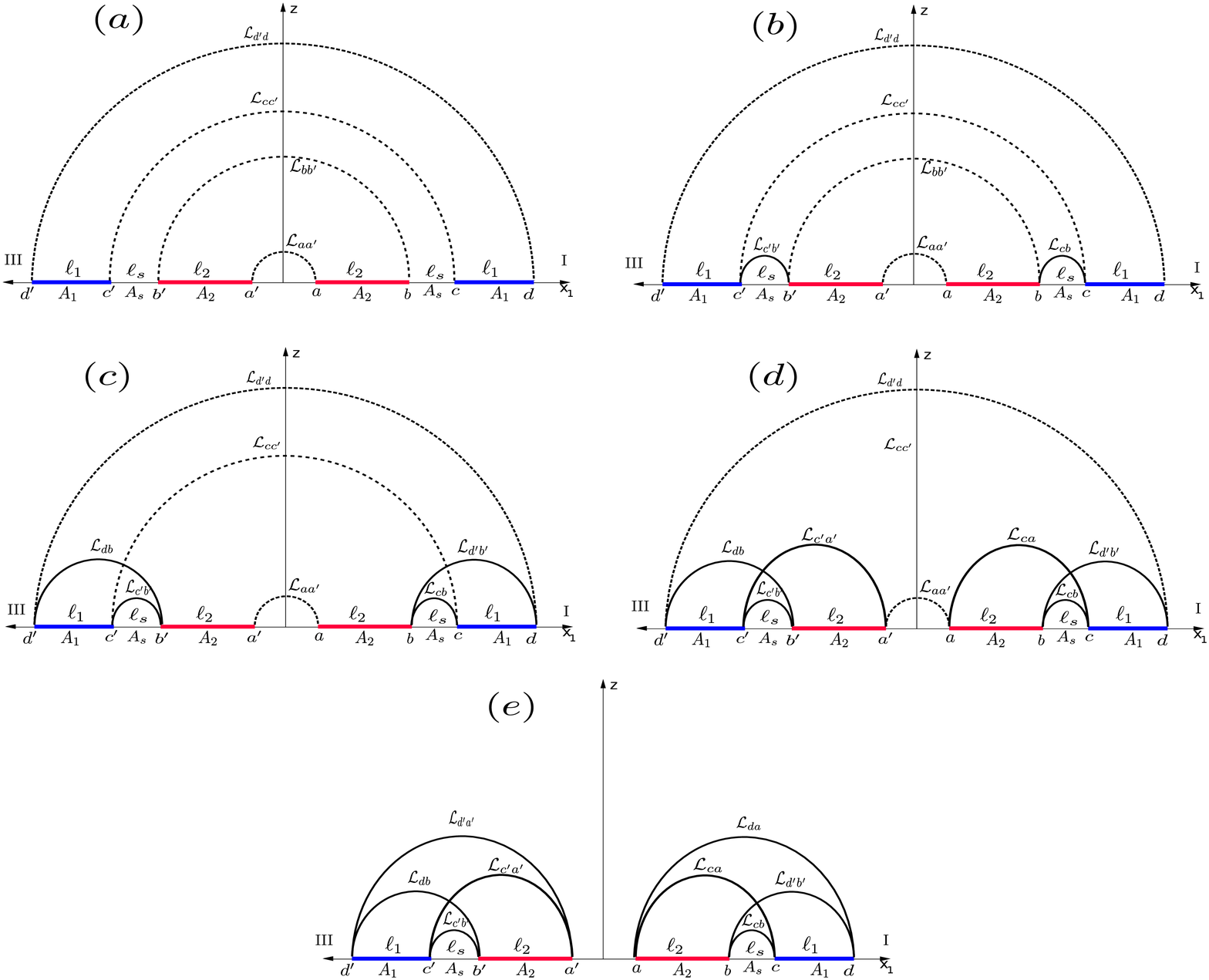}
\caption{Geodesic configurations for differnt limits of the time relative to the lengths of the intervals. The geodesics depicted by dashed lines pass through the interior region of the BTZ black string connecting the two Rindler wedges $I$ and $III$. The geodesics depicted by solid lines are confined to the exterior region of the BTZ black string outside the horizon. }\label{neggeodjfin}
\end{figure}

In this limit the geodesics (solid curves) that joins $c'$ and $b'$ and $c$ and $b$ are confined to the exterior region, and rest of the geodesics (dashed curves) pass through the interior region of the BTZ black string as depicted in Fig. \ref{neggeodjfin}(b). The holographic entanglement negativity in this case may then be obtained by utilizing both the eqs. \eqref{earlytimes} and \eqref{latetimes} in the eq. \eqref{eq_holo_ent_neg_dj_int_holo_ent_enpy}, which leads to the following expression
\begin{equation}\label{djlimb}
\begin{aligned}
\mathcal{E} = &\frac{3}{16G^{(3)}_N}\left(\mathcal{L}_{d'd}+\mathcal{L}_{b'b}+\mathcal{L}_{c'c}+\mathcal{L}_{a'a}-\mathcal{L}_{d'd}-\mathcal{L}_{a'a}-\mathcal{L}_{c'b'}-\mathcal{L}_{cb}\right),\\
=&\frac{3}{16G^{(3)}_N}\left(\mathcal{L}_{b'b}+\mathcal{L}_{c'c}-\mathcal{L}_{c'b'}-\mathcal{L}_{cb}\right),\\
=&\frac{\pi c}{\beta}\left(t-\frac{l_s}{2}\right).
\end{aligned}
\end{equation}
\subsubsection{$\ell_2/2<(\ell_1+\ell_s)/2<t<(\ell_2+\ell_s)/2$}

In this case the geodesics (solid curves) between the points $d'$ to $b'$, $d$ to $b$, $c'$ to $b'$ and $c$ to $b$ are confined in the exterior region  while the other geodesics (dashed curves) between the points $d'$ to $d$ and $c'$ to $c$ pass through the interior region of the BTZ black string as depicted in Fig. \ref{neggeodjfin}(c). The holographic entanglement negativity in this limit is obtained by utilizing eqs. \eqref{earlytimes} and \eqref{latetimes} in the eq. \eqref{eq_holo_ent_neg_dj_int_holo_ent_enpy} leading to the following expression
\begin{equation}\label{djlimc}
\begin{aligned}
\mathcal{E} = &\frac{3}{16G^{(3)}_N}\left(\mathcal{L}_{d'b'}+\mathcal{L}_{db}+\mathcal{L}_{c'c}+\mathcal{L}_{a'a}-\mathcal{L}_{d'd}-\mathcal{L}_{a'a}-\mathcal{L}_{c'b'}-\mathcal{L}_{cb}\right),\\
=&\frac{3}{16G^{(3)}_N}\left(\mathcal{L}_{d'b'}+\mathcal{L}_{db}+\mathcal{L}_{c'c}-\mathcal{L}_{d'd}-\mathcal{L}_{c'b'}-\mathcal{L}_{cb}\right),\\
=&\frac{\pi c\ell_1}{2\beta}.
\end{aligned}
\end{equation}
\subsubsection{$(\ell_1+\ell_s)/2<(\ell_2+\ell_s)/2<t<(\ell_1+\ell_2+\ell_s)/2$}

In this limit the geodesics (dashed curves) between the points $d'$ to $d$ and $a'$ to $a$ pass through the interior region whereas rest of the geodesics (solid curves) are confined to the exterior region of the BTZ black string. The holographic entanglement negativity in this case may be obtained by utilizing eqs. \eqref{earlytimes} and \eqref{latetimes} in the eq. \eqref{eq_holo_ent_neg_dj_int_holo_ent_enpy}, as follows
\begin{equation}\label{djlimd}
\begin{aligned}
\mathcal{E} = &\frac{3}{16G^{(3)}_N}\left(\mathcal{L}_{d'b'}+\mathcal{L}_{db}+\mathcal{L}_{c'a'}+\mathcal{L}_{ca}-\mathcal{L}_{d'd}-\mathcal{L}_{a'a}-\mathcal{L}_{c'b'}-\mathcal{L}_{cb}\right),\\
=&\frac{\pi c}{\beta}\left(\frac{\ell_1+\ell_2+\ell_s}{2}-t \right)\,.
\end{aligned}
\end{equation}
\subsubsection{$(\ell_1+\ell_s)/2<(\ell_2+\ell_s)/2<(\ell_1+\ell_2+\ell_s)/2<t$}

In the late time limit all the geodesics (solid curves) are confined to the exterior region  of the BTZ black string as shown in Fig. \ref{neggeodjfin}(e). Consequently, the holographic entanglement negativity is obtained by utilizing  eqs. \eqref{earlytimes} and \eqref{latetimes} in the eq. \eqref{eq_holo_ent_neg_dj_int_holo_ent_enpy} which vanishes due to mutual cancellations as follows
\begin{equation}\label{djlime}
\begin{aligned}
\mathcal{E} = &\frac{3}{16G^{(3)}_N}\left(\mathcal{L}_{d'b'}+\mathcal{L}_{db}+\mathcal{L}_{c'a'}+\mathcal{L}_{ca}-\mathcal{L}_{d'a'}-\mathcal{L}_{da}-\mathcal{L}_{c'b'}-\mathcal{L}_{cb}\right),\\
=&0.
\end{aligned}
\end{equation}

It is now possible to present a holographic description for the time evolution of the entanglement negativity for mixed state configurations of two disjoint intervals in a $CFT_{1+1}$ following a global quench. As described earlier the corresponding bulk dual geometry is the single sided black hole obtained by slicing the eternal black hole geometry in half with an ETW brane. So the time evolution of the entanglement negativity for the mixed state in question following a global quench may be obtained by considering half of the corresponding results for the eternal black hole geometry computed above.
Quite significantly, the time evolution of the holographic entanglement negativity obtained from our construction in this case matches exactly with the corresponding replica technique results for a $CFT_{1+1}$ as described in section (\ref{ent_neg_glo_quench_dj_int}). Once again this is a robust consistency check for our holographic construction.

\begin{figure}[H]
\centering
\includegraphics[width=.5\linewidth]{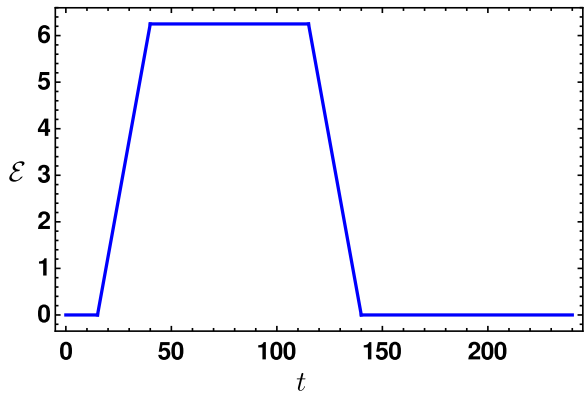}
\caption{Holographic entanglement negativity $\mathcal{E}$ vs time $t$ is plotted for two disjoint intervals of lengths $\ell_1=50$, $\ell_2=200$ with the seperation length $l_s=30$.}\label{neg_vs_t_dj_int}
\end{figure}

In  Fig. \ref{neg_vs_t_dj_int} we have plotted the time evolution of the holographic entanglement negativity for the mixed state in question in a $CFT_{1+1}$ following a global quench.  It is observed from Fig. \ref{neg_vs_t_dj_int} that the holographic entanglement negativity for early times is zero and grows linearly to attain a maximum value at an intermediate time after which remains constant upto certain value of time. Beyond this value of time the holographic entanglement negativity decreases linearly and vanishes for late times. Note that this behavior for the entanglement negativity of the mixed state in question consistently matches with that described for a $CFT_{1+1}$s as described in \cite{Coser:2014gsa}.
\subsection{Two adjacent intervals}\label{henadjagq}

\begin{figure}[h]
\centering
\includegraphics[width=\linewidth]{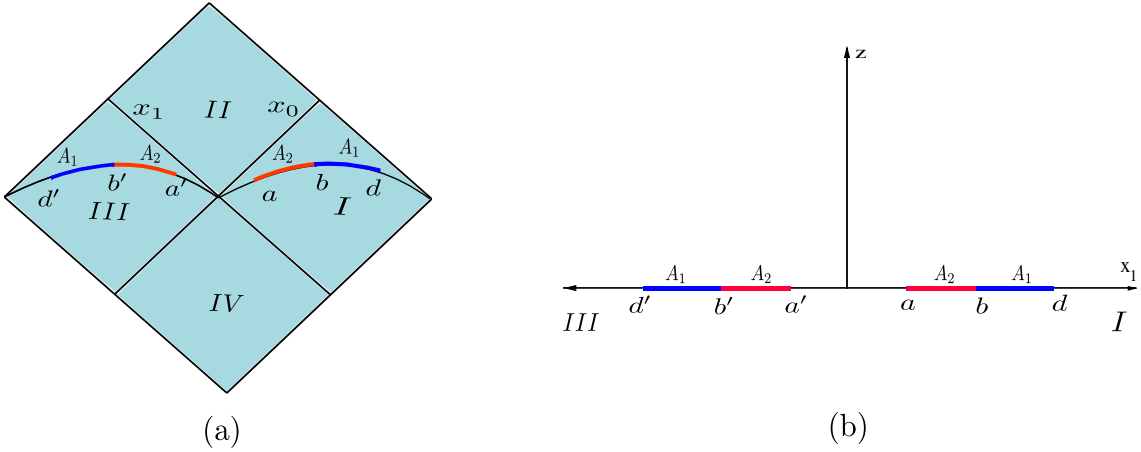}
\caption{(a) Two adjacent intervals $A_1$ (blue) and $A_2$ (red) in the Rindler wedges $I$ and $III$ respectively of the Minkowski diamond. (b) These two intervals are projected on the $x_1$-$z$ plane of the Poincar\'e coordinates.}\label{coord_adj_intervals}
\end{figure}

We now consider the time evolution of the entanglement negativity for the case of two adjacent intervals $A_1$ and $A_2$ of lengths $\ell_1$ and $\ell_2$ respectively. Proceeding similarly as in the case for the two disjoint intervals described above, we consider the two adjacent intervals $A_1$ and $A_2$ in both the Rindler wedges $I$ and $III$ of the Minkowski diamond as shown in Fig. \ref{coord_adj_intervals}(a). The end points of the intervals $A_1$ and $A_2$ in the Rindler wedges projected on the $x_1 -z$ plane of the Poincaré coordinates are denoted as ($a, b$), ($b, c$) and ($a', b'$), ($b', c'$)  respectively which is depicted in Fig. \ref{coord_adj_intervals}(b).

Following the analysis described earlier in the last subsection for the mixed state of disjoint intervals it is also possible to obtain the time evolution of the entanglement negativity for the case of two adjacent intervals under consideration here. Once again for this purpose it is required to consider the dual bulk eternal black hole geometry and utilize eqs. \eqref{heecon}, \eqref{earlytimes} and \eqref{latetimes} to obtain the corresponding holographic entanglement negativity. In this case also distinct sets of geodesics contribute to the holographic entanglement negativity for various values of the time $t$ relative to the lengths of the intervals $\ell_1$ and $\ell_2$. As earlier we consider these interesting limits and elucidate the corresponding geodesic structures leading to the holographic entanglement negativity for each of these cases.

\subsubsection{$t<\ell_1/2<\ell_2/2<(\ell_1+\ell_2)/2$}
In the limit of early times all of the geodesics pass through the interior region of the BTZ black string connecting the one Rindler wedge $I$ to the other $III$,  depicted by dashed curves in the Fig. \ref{neggeoadjfin}(a). The corresponding holographic entanglement negativity is obtained by employing eq. \eqref{earlytimes} in the eq. \eqref{heecon}, which is given as follows
\begin{equation}\label{adjlima}
\begin{aligned}
\mathcal{E} = &\frac{3}{16G^3_N}(\mathcal{L}_{c'c}+\mathcal{L}_{b'b}+\mathcal{L}_{b'b}+\mathcal{L}_{a'a}-\mathcal{L}_{c'c}-\mathcal{L}_{a'a}),\\
=&2\mathcal{L}_{b'b}=\frac{\pi ct}{\beta}.
\end{aligned}
\end{equation}

\subsubsection{$\ell_1/2<t<\ell_2/2<(\ell_1+\ell_2)/2$}

\begin{figure}[]
\centering
\includegraphics[width=\linewidth]{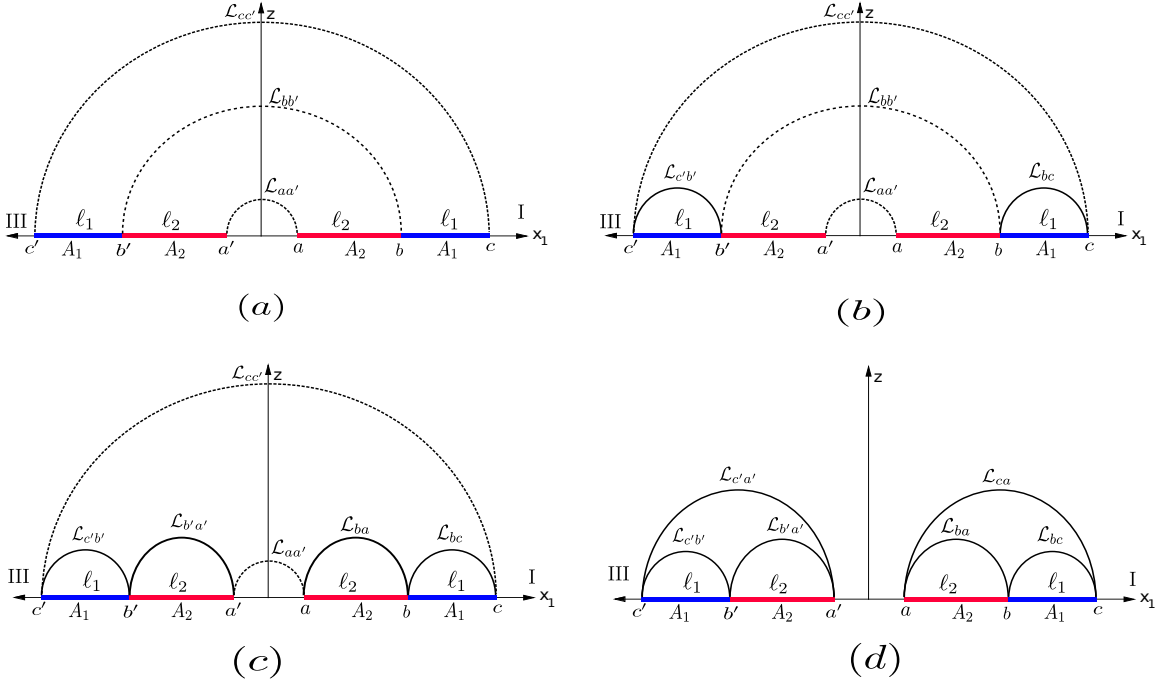}
\caption{Geodesic configurations for differnt limits of the time with the lenghts of the intervals. The geodesics shown by dashed line pass through the interior region of the BTZ black hole connecting two rindler wedges. The geodesics shown by solid line confine at the exterior region of the BTZ black hole and they do not pass through the interior region.}\label{neggeoadjfin}
\end{figure}

In this limit the geodesics (solid curves) between the points $c'$ to $b'$ and $c$ to $b$ are confined in the exterior region, rest of the geodesics pass through the interior region of the BTZ black string connecting the Rindler wedge $I$ to the other $III$ as shown in Fig. \ref{neggeoadjfin}(b). The holographic entanglement negativity in this limits is given as
\begin{equation}\label{adjlimb}
\begin{aligned}
\mathcal{E} = &\frac{3}{16G^3_N}\left(\mathcal{L}_{c'b'}+\mathcal{L}_{cb}+\mathcal{L}_{b'b}+\mathcal{L}_{a'a}-\mathcal{L}_{c'c}-\mathcal{L}_{a'a}\right),\\
=&\left(\mathcal{L}_{c'b'}+\mathcal{L}_{cb}+\mathcal{L}_{b'b}-\mathcal{L}_{c'c}\right)=\frac{\pi c\ell_1}{2\beta}.
\end{aligned}
\end{equation}

\subsubsection{$\ell_1/2<\ell_2/2<t<(\ell_1+\ell_2)/2$}

In this case only the geodesics (dashed curves) between the points $c'$ to $c$ and $c$ to $b$ pass through the interior region and rest of the geodesics remain in the exterior region of the BTZ black string as depicted in Fig. \ref{neggeoadjfin}(c). The holographic entanglement negativity in this case is given as follows
\begin{equation}\label{adjlimbc}
\begin{aligned}
\mathcal{E} = &\frac{3}{16G^3_N}\left(\mathcal{L}_{c'b'}+\mathcal{L}_{cb}+\mathcal{L}_{b'a'}+\mathcal{L}_{ba}-\mathcal{L}_{c'c}-\mathcal{L}_{a'a}\right),\\
=&\frac{\pi c}{\beta}\left(\frac{\ell_1+\ell_2}{2}-t \right).
\end{aligned}
\end{equation}

\subsubsection{$\ell_1/2<\ell_2/2<(\ell_1+\ell_2)/2<t$}

In this limit all the geodesics (solid curves)  are confined in the exterior region of the BTZ black string as depicted in Fig. \ref{neggeoadjfin}(d). The resulting holographic entanglement negativity vanishes due to mutual cancellations as follows 
\begin{equation}\label{adjlimd}
\begin{aligned}
\mathcal{E} = &\frac{3}{16G^3_N}\left(\mathcal{L}_{c'b'}+\mathcal{L}_{cb}+\mathcal{L}_{b'a'}+\mathcal{L}_{ba}-\mathcal{L}_{c'a'}-\mathcal{L}_{ca}\right),\\
=&0
\end{aligned}
\end{equation}

As mentioned earlier the time evolution of the holographic entanglement negativity for two adjacent intervals after a global quench in a $CFT_{1+1}$ dual to a single sided black hole is obtained by taking half of the corresponding results for the eternal black hole case. Interestingly, it is observed that the time evolution of the holographic entanglement negativity in question following a global quench match exactly with the corresponding $CFT_{1+1}$ results computed through the replica technique in eq. (\ref{ent_neg_glo_quench_adj_int}).
\begin{figure}[H]
\centering
\includegraphics[width=.5\linewidth]{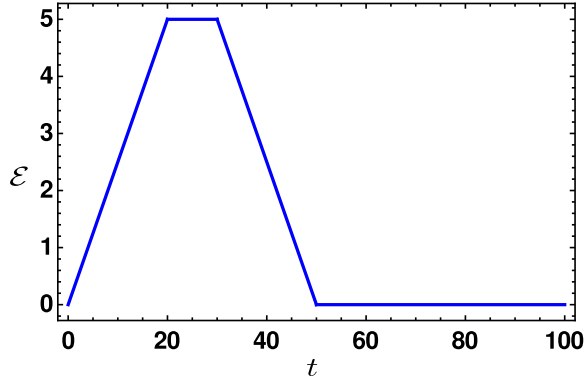}
\caption{Holographic entanglement negativity $\mathcal{E}$ vs time $t$ is plotted for two adjacent intervals of lengths $\ell_1=50$, $\ell_2=200$.}\label{plotajdintneg}
\end{figure}

The time evolution of the holographic entanglement negativity is illustrated in Fig. \ref{plotajdintneg}. The holographic entanglement negativity grows linearly for the initial part of its evolution and reaches to its maximum value at an intermediate time beyond which it is a constant upto a certain range of time. Beyond this range the holographic entanglement negativity decreases linearly and becomes zero for late times. Note that this behavior of the holographic entanglement negativity is consistent with that described in \cite{Coser:2014gsa}.

\section{Summary and discussion}\label{sumanddis}

To summarize, in this article we have investigated the time evolution of the entanglement negativity for mixed states of two disjoint and adjacent intervals following a global quench in a holographic $CFT_{1+1}$ through the $AdS_3/CFT_2$ correspondence. In this context we have considered the holographic dual to the global quench scenario as the two sided eternal black hole sliced in half by an end of the world (ETW) brane, where the initial state is the conformal invariant boundary state (B-state). The corresponding holographic entanglement negativity is given by half of the result for the eternal black hole geometry. For our holographic construction these intervals were considered on both the Rindler wedges of the Minkowski diamond and projected on the $x_1-z$ plane of the 
Poincar\'e coordinates for the dual eternal black hole geometry. 

Our holographic construction follows from the replica technique observation that the universal parts of corresponding four point and the three point twist correlators on a strip in the $CFT_{1+1}$, which are related to the entanglement negativity for these mixed states, admit a factorization in terms of the product of certain two point twist correlators on a strip. The entanglement negativity may then be expressed as a specific algebraic sum of the entanglement entropies for the intervals and certain appropriate combinations of them. It is then possible to employ the Hartman-Maldacena prescription for the holographic entanglement entropy 
in a dual eternal black geometry from the appropriate geodesic combinations to obtain the holographic entanglement negativity
of the mixed states in question. The corresponding holographic entanglement negativity for these mixed states in a $CFT_{1+1}$ after a global quench may then be computed by considering half of the above results for the eternal black hole geometry (which describes a single sided black hole with an ETW brane).

For our case the dual bulk configuration is described by the BTZ black string and it is required to consider the combination of geodesics located exterior to the horizon and those passing through the interior region connecting the two Rindler wedges of the Minkowski diamond. Interestingly it is observed that 
the holographic entanglement negativity at early times receive contributions from the lengths of the geodesics passing through the interior region of the bulk BTZ black string. For late times on the other hand the contribution to the holographic entanglement negativity arise from the geodesics restricted to the region exterior to the horizon in both the Rindler wedges.

The corresponding time evolution of the entanglement negativity following a global quench for the disjoint and the adjacent intervals, obtained from half the result for that of the bulk eternal black hole, exhibits a similar behavior. For early times the entanglement negativity is zero and increases linearly to reach a saturation at an intermediate value and remains constant upto a certain range of time. Beyond this range the entanglement negativity decreases linearly with time and vanishes for late times which corresponds to the phenomena of the $CFT$ thermalization indicating the formation of a black hole in the bulk. Interestingly, the holographic entanglement negativity obtained by us match exactly with the corresponding $CFT_{1+1}$ replica technique results providing a robust consistency check for our construction. Our analysis described in this article provides an interesting and significant insight into the 
structure and evolution of entanglement negativity for mixed states in conformal field theories following a global quench from a holographic perspective and indicates a possible higher dimensional extension to a generic $AdS_{d+1}/CFT_{d}$ scenario. However such a higher dimensional extension of our construction will require certain substantiation through explicit examples for relevant consistency checks. We hope to return to these and other related interesting issues in the near future.


\bibliographystyle{JHEP}
\bibliography{HENGQ}

\end{document}